\documentclass[12pt]{article}

\ifx\pdfoutput\undefined
\usepackage[dvips,bookmarks]{hyperref}
\else
\usepackage{hyperref}
\fi
\hypersetup{colorlinks=false,bookmarksopen,bookmarksnumbered,citecolor=blue,
   pdfstartview=FitH}

\usepackage{latexsym}
\usepackage{amssymb,amsfonts,amsmath}
\usepackage{graphicx} 
\usepackage{indentfirst}
\usepackage{bbm}
\usepackage{amssymb}
\usepackage{verbatim}
\usepackage{amsmath, amsthm,amssymb}
\usepackage{mathrsfs}
\usepackage{hyperref}
\usepackage{amsfonts}
\usepackage{dsfont}
\parskip=\medskipamount

\newcommand {\cD}{{\cal D}}
\newcommand {\cE}{{\cal E}}
\newcommand {\cF}{{\cal F}}
\newcommand {\cG}{{\cal G}}

\newcommand {\cM}{{\cal M}}
\newcommand {\cN}{{\cal N}}

\newcommand {\cP}{{\cal P}}

\newcommand {\cR}{{\cal R}}

\def\a{\alpha}
\def\b{\beta}
\def\c{\chi}
\def\d{\delta}

\def\f{\phi}
\def\g{\gamma}
\def\G{\Gamma}

\def\m{\mu}

\def\o{\omega}
\def\p{\pi}
\def\q{\theta}

\def\s{\sigma}

\def\D{\Delta}
\def\F{\Phi}
\def\J{\Psi}
\def\L{\Lambda}
\def\O{\Omega}

\def\X{\Xi}
\def\tr{{\rm tr}}

\def\ri{{\rm i}}
\def\re{{\rm e}}

\newcommand{\ad}{{\dot{\alpha}}}                           %new
\newcommand{\bd}{{\dot{\beta}}}                            %new
\newcommand{\ve}{\varepsilon}                            %new
\newcommand{\cDB}{{\bar\cD}}                            %new

\newcommand{\pa}{\partial}                           %new
\newcommand{\hf}{\frac12}
\newcommand{\vf}{\varphi}
\newcommand{\be}{\begin{equation}}
\newcommand{\ee}{\end{equation}}
\newcommand{\bea}{\begin{eqnarray}}
\newcommand{\eea}{\end{eqnarray}}
\newcommand{\non}{\nonumber}
\newcommand{\ba}{\begin{array}}
\newcommand{\ea}{\end{array}}

\def\double #1{#1{\hbox{\kern-2pt $#1$}}}

\newcommand{\gd}{{\dot\g}}

\newcommand{\bsubeq}{\begin{subequations}}
\newcommand{\esubeq}{\end{subequations}}

\newcommand{\rd}{\mathrm d}
\numberwithin{equation}{section}
\begin{document}
%%%%%%%%%%%%%%%%
%%%%%%%%%%%%%%%%

\begin{titlepage}
\begin{flushright}
Nikhef-2013-022\\
July, 2013\\
\end{flushright}
\vspace{5mm}

\begin{center}
{\Large \bf 
Nonlocal action for the super-Weyl anomalies: \\
A new representation}
\\ 
\end{center}

\begin{center}

{\bf
Daniel Butter${}^{a}$ and Sergei M. Kuzenko${}^{b}$} \\
\vspace{5mm}

\footnotesize{
${}^{a}${\it Nikhef Theory Group \\
Science Park 105, 1098 XG Amsterdam, The Netherlands}}
~\\
\texttt{dbutter@nikhef.nl}\\
\vspace{2mm}

\footnotesize{
${}^{b}${\it School of Physics M013, The University of Western Australia\\
35 Stirling Highway, Crawley W.A. 6009, Australia}}  
~\\
\vspace{2mm}

\end{center}

\begin{abstract}
\baselineskip=14pt
Using the recently discovered $\cN=1$ supersymmetric extension 
of the conformal fourth-order scalar operator (introduced
originally by Fradkin and Tseytlin and also known as the
``Paneitz operator'' or ``Riegert operator''), 
we derive a new representation for the nonlocal action generating 
the super-Weyl anomalies. 
\end{abstract}

\vfill

\vfill
\end{titlepage}

\newpage
\renewcommand{\thefootnote}{\arabic{footnote}}
\setcounter{footnote}{0}

%\tableofcontents

%\newpage

%%%%%%%%%%%%%%%%%%%%%%%%%%%%%%%%%%%%%%%%%%%%%%%%%%%%%%
%%%%%%%%%%%%%%%%%%%%%%%%%%%%%%%%%%%%%%%%%%%%%%%%%%%%%%

\section{Introduction}

The concept of super-Weyl transformations \cite{HT, Siegel:SC} in the framework of 
the Wess-Zumino formulation \cite{WZ,GWZ} for 4D $\cN=1$ supergravity 
is a generalization of the ordinary Weyl transformations.
The property of a locally supersymmetric theory to be super-Weyl invariant 
can naturally be recast in terms of the Ferrara-Zumino supercurrent multiplet \cite{FZ} 
which contains the energy-momentum tensor and the spinor supersymmetry current along 
with some other components.  It is pertinent for our subsequent discussion
to elaborate on this point in some more detail.

The Ferrara-Zumino  multiplet is realized  in terms of two constrained superfields:  
a real vector  $T_{a} $ known as the supercurrent  and  a covariantly chiral 
scalar $T$,  $\bar \cD_\ad T =0$, called the supertrace
(also known as the anomaly multiplet).
The supercurrent is the source of supergravity  \cite{OS,FZ78,Siegel77}. 
Let $S [\c, H, \vf , \bar \vf] $ be the action of matter superfields $\c^i$ coupled 
to background supergravity. The supergravity multiplet is fully described in terms 
of the corresponding prepotentials $H^m$, $\vf$ and $\bar \vf$ \cite{Siegel78,SG}, 
of which the gravitational superfield \cite{OS} $H^m$ is real and the
conformal compensator \cite{Siegel78} $\vf$ is chiral.   
The supercurrent and the supertrace originate as variational derivatives of the matter action 
with respect to the supergravity prepotential. Specifically, it turns out that 
\begin{subequations} \label{1.1}
\bea
T_a &=& \frac{\D } {\D H^a} S [\c, H, \vf , \bar \vf]~, \qquad \\
T &=& \frac{\d } {\d \s} S [\c, H, \re^\s \vf , \re^{\bar \s} \bar \vf]\Big|_{\s=\bar \s=0}~, 
\qquad \bar \cD_\ad \s=0~,
\label{supertrace}
\eea
\end{subequations}
where $\s$ is an arbitrary covariantly chiral scalar
superfield. 
Here $\D/ \D H^{a}$ denotes a covariantized variational derivative with respect to the gravitational superfield, see \cite{GGRS,BK} for pedagogical reviews. 
If the matter fields are chosen to obey their equations of motion, $\d S/ \d \c^i =0$, 
the condition that the matter action is locally supersymmetric is expressed 
as the conservation equation
\bea
\bar \cD^\ad T_{\a\ad} +\frac{2}{3} \cD_\a T = 0~.
\eea

A super-Weyl transformation \cite{HT} is  a local rescaling of the chiral 
compensator \cite{Siegel:SC}, $\vf \to \vf' = \re^\s \vf$, accompanied by a certain transformation 
of the matter superfields of the form $\c^i \to \c'{}^i = \re^{-d_{(+)} \s - d_{(-)} \bar \s}\c^i$, 
for some parameters $d_{(\pm)} $
(with the gravitational superfield $H^m$ being super-Weyl inert). 
If the matter fields in \eqref{1.1} are chosen to satisfy 
the equations of motion $\d S/ \d \c^i =0$, it follows from \eqref{supertrace}
that $T$ determines the super-Weyl variation of the action. 
Without imposing the matter equations of motion, 
the super-Weyl variation of the action is
\bea
\d_\s S =  
\int \rd^4 x \,\rd^2 \q \,\cE \, \s T ~+ ~{\rm c.c.} ~+~\int \d_\s\c^i  \cdot \frac{\d S}{\d \c^i} ~,
\eea 
where $\cE= \vf^3$ denotes the chiral integration measure 
(see, e.g.,  \cite{GGRS,BK} for more details).
If the theory is super-Weyl invariant, then $T=0$ for the on-shell matter fields. 
In a rigid supersymmetric limit, when $H^m \to 0$ and $\vf \to 1$, 
such a theory becomes superconformal. 

In the quantum theory, integrating out the matter fields leads to the 
effective action $\G  [ H, \vf , \bar \vf]$. The quantum supertrace is
\bea
\langle T\rangle =  \frac{\d } {\d \s} \G [ H, \re^\s \vf , \re^{\bar \s} \bar \vf]\Big|_{\s=\bar \s=0}~.
\label{1.4}
\eea
Generically, the super-Weyl symmetry is anomalous at the quantum level. 
This means that, if one starts from a classically super-Weyl invariant theory, 
then in general $\langle T\rangle $ turns out to be a non-zero local functional 
of the supergravity prepotentials, which cannot  be eliminated by adding local
counterterms to the effective action.
According to the cohomological analysis given in \cite{BPT}, 
the general  form of $\langle T\rangle $
in classically super-Weyl invariant theories  is
\bea
8\p^2 \langle T\rangle= c \,W^{\a\b\g} W_{\a\b\g} - a \,\cG 
 +\frac{1}{16}h\, (\bar \cD^2 - 4R ) \cD^2 R ~,
\label{anomaly}
\eea
where $\cG$ denotes the following chiral scalar:
\bea
\cG =  W^{\a\b\g} W_{\a\b\g} -\frac{1}{4} (\bar \cD^2 - 4R ) (G^aG_a +2R \bar R)~.
\eea
It turns out that the functional  \cite{FZ78} (with $E$ the full superspace integration measure)
\bea
\int \rd^4 x \,\rd^2 \q \,\cE \, \cG = \int \rd^4 x \,\rd^2 \q \,\cE \,W^{\a\b\g} W_{\a\b\g} 
+ \int \rd^4 x \,\rd^2 \q \,\rd^2 \bar \q \,E \, (G^aG_a +2R \bar R)
\eea
is a topological invariant (see  \cite{Buchbinder:1988tj,BK} 
for the  proof),  which is related to the difference 
of the Gauss-Bonnet and Pontryagin invariants. 
The coefficients $a$ and $c$ in \eqref{anomaly}
are of nontrivial significance in general superconformal 
field theories, see \cite{AFGJ,Osborn,SchT,KSch} and references therein.
On the other hand, the coefficient $h$ may be completely arbitrary, for its value may be freely changed
by adding to the effective action 
a finite local counterterm proportional to 
\bea
\int \rd^4 x \,\rd^2 \q \,\rd^2 \bar \q \,E \, R\bar R~.
\label{RbarR}
\eea

The explicit calculation of the anomalous supertrace \eqref{anomaly} 
for  the chiral and vector multiplets was carried out in \cite{BK86} 
(see \cite{BK} for a review).
This calculation made use of the super-$a_2$ (or, equivalently, super-$b_4$)
coefficients computed by McArthur \cite{McA2} (for the chiral multiplet)
using the superspace normal coordinates
\cite{McA} and independently in \cite{BK86} using the superfield Schwinger-DeWitt 
technique. Given the anomalous supertrace  \eqref{anomaly}, it becomes possible to look for
a nonlocal effective action that generates the anomaly.
Such an action was derived in \cite{BK88} (see \cite{BK} for a review).  
This was done with the aid of 
a composite chiral scalar $\O$, introduced originally in \cite{GNSZ},
which obeys the massless equation 
\bea
( \cD^2 -4\bar R ) \O=0~, \qquad \bar \cD_\ad \O =0
\eea
and possesses the super-Weyl transformation law\footnote{Using the super-Weyl transformation 
of the antichiral torsion $\bar R$, 
$\bar R \to \bar R' = -\frac{1}{4} \re^{-2\bar \s} 
(\cD^2 -4 \bar R) \re^\s$, one can see that $\O$ defines a super-Weyl transformation 
such that $\bar R' = 0$ and $R'=0$.} 
\bea
\O \to \O' = \re^{-\s} \O~.
\eea
The explicit expression for $\O$ is
\bea
\O = 1+ \frac{1}{4 \Box_+} ( \bar \cD^2 -4R) \bar R, 
\eea
where $\Box_+$ denotes the chiral d'Alembertian defined by 
$\Box_+ \f = \frac{1}{16} ( \bar \cD^2 -4R) ( \cD^2 -4 \bar R) \f$, 
for any covariantly chiral scalar $\f$. 

Conceptually, the chiral superfield $\O$ is a supersymmetric analogue 
of the composite scalar field $\o= 1 + \frac{1}{6} (\Box - \frac{1}{6} \cR)^{-1} \cR$,  
with $\cR$ the scalar curvature,
introduced by Fradkin and Vilkovisky \cite{FV}.    The scalar field $\o$ was used
by Fradkin and Tseytlin \cite{FT84} to integrate the ordinary Weyl (or conformal) anomalies
(see \cite{DDI,Duff,DSch,Duff94} and references therein).
So the approach of \cite{BK88} may be thought of as a supersymmetric extension 
of that given in \cite{FT84}. In the case of the Weyl anomaly, 
there exists an alternative method \cite{Riegert}
to construct a nonlocal action generating the anomaly.\footnote{Nonlocal actions
for Weyl anomalies in  higher dimensions 
were studied in \cite{KMM,Deser96,Deser2000}.} 
It makes use of the conformal fourth-order scalar operator 
\begin{align} \label{D_0}
\Delta_0 = \Box \Box +  \mathcal{D}^a \big(
	2 \mathcal{R}_{ab} \,\mathcal{D}^b 
	- \tfrac{2}{3} \mathcal{R} \,\mathcal{D}_a
	\big)
\end{align}
discovered by Fradkin and Tseytlin\footnote{Fradkin and
Tseytlin \cite{FT1982} also constructed conformal
operators $\Delta_s$ for fields of different spin.} 
in 1981 \cite{FT1982}
and re-discovered 
by Paneitz in 1983 \cite{Paneitz}.\footnote{In the mathematics
literature, this operator is known as the Paneitz operator. 
In the first preprint version of our paper, it was called 
``the Paneitz-Riegert operator'' since the same operator had appeared
in \cite{Riegert}  and we were not aware of the earlier work by Fradkin and Tseytlin
\cite{FT1982}.} 

It was mentioned in \cite{BK88} that the technique of \cite{Riegert} was not ``directly applicable 
for the anomaly integration in curved superspace.'' The reason for that was the non-existence 
of a super-Weyl covariant 
chiral d'Alembertian\footnote{Given a chiral  d'Alembertian $\stackrel{\frown}{\Box}_+$ such that 
$\stackrel{\frown}{\Box}_+ \J$ is covariantly chiral for any covariantly chiral scalar  $\J$, 
the differential part of $\stackrel{\frown}{\Box}_+$ is uniquely fixed \cite{BK86},
$\stackrel{\frown}{\Box}_+ = \cD^a\cD_a +\frac{1}{4}R\cD^2 +\ri G^a \cD_a +\frac{1}{4}(\cD^\a R)\cD_\a +\cP$,  where the scalar superfield $\cP$ is covariantly chiral. The only free parameter in $\stackrel{\frown}{\Box}_+ $
is the chiral scalar $\cP$.}
$\stackrel{\frown}{\Box}_+$ such that the functional
\bea
\int \rd^4 x \,\rd^2 \q \,\rd^2 \bar \q \,E \, \bar \F \stackrel{\frown}{\Box}_+ \J
=\frac{1}{16} \int \rd^4 x \,\rd^2 \q \,\rd^2 \bar \q \,E \,  (\bar \cD^2 \bar \F ) \cD^2 \J + \dots
\label{1.11}
\eea
is super-Weyl invariant for any super-Weyl inert chiral scalars $\F$ and $\J$
(the ellipsis on the right of \eqref{1.11} denotes terms with two or fewer spinor derivatives).
The component counterpart of the operator in \eqref{1.11} is fourth-order in
vector derivatives.

A way out has only recently been found in
Ref. \cite{BdeWKL} which provided the required supersymmetric extension of 
the Fradkin-Tseytlin $\Delta_0$ operator (using the novel 
superspace formulation \cite{ButterN=1}  for $\cN=1$ conformal  supergravity).
Here we will use this operator to 
derive a new representation for the nonlocal action generating 
the super-Weyl anomaly.

 This paper is organized as follows. In section 2,  we describe the properties 
 of the supersymmetric Fradkin-Tseytlin operator. In section 3, 
 we derive the nonlocal effective action generating the super-Weyl anomaly. 
Concluding comments are given in section 4. 
 A brief summary of the Wess-Zumino superspace geometry, 
following the notation and conventions of \cite{BK}, is given in the appendix.

\section{The supersymmetric Fradkin-Tseytlin operator}

In this section we give a detailed derivation of the $\cN=1$ super-Weyl covariant 
operator which was introduced in \cite{BdeWKL} as the 
supersymmetric extension of the Fradkin-Tseytlin  operator \eqref{D_0}.

As is well known, the algebra of covariant derivatives \eqref{algebra} 
is invariant under the infinitesimal super-Weyl transformation \cite{HT} 
associated with a chiral parameter $\s$, $\bar \cD_\ad \s=0$, and its complex conjugate $\bar \s$,
\begin{subequations} 
\label{superweyl}
\bea
\d_\s \cD_\a &=& ( \hf \s - {\bar \s} )  \cD_\a - (\cD^\b \s) \, M_{\a \b}  ~, \\
\d_\s \bar \cD_\ad & = & ( \hf {\bar \s} - \s  )
\bar \cD_\ad -  ( \bar \cD^\bd  {\bar \s} )  {\bar M}_{\ad \bd} ~,\\
\d_\s \cD_{\a\ad} &=& -\hf( \s +\bar \s) \cD_{\a\ad} 
-\frac{\ri}{2} (\bar \cD_\ad \bar \s) \cD_\a - \frac{\ri}{2} ( \cD_\a  \s) \bar \cD_\ad \non \\
&& - (\cD^\b{}_\ad \s) M_{\a\b} - (\cD_\a{}^\bd \bar \s) \bar M_{\ad \bd}~,
\eea
\end{subequations}
provided the torsion tensors transform\footnote{The super-Weyl transformation of $G_{\a\ad}$
given in \cite{BK}, eq. (5.5.14), contains a typo.}
 as follows:
\begin{subequations} 
\label{superweyl-torsion}
\bea
\d_\s R &=& -2\s R -\frac{1}{4} (\bar \cD^2 -4R ) \bar \s 
= (\bar \s - 2\s) R -\frac{1}{4} \bar \cD^2 \bar \s
~, \\
\d_\s G_{\a\ad} &=& -\hf (\s +\bar \s) G_{\a\ad} +\ri \,\cD_{\a\ad} (\bar \s- \s) ~, 
\label{s-WeylG}\\
\d_\s W_{\a\b\g} &=&-\frac{3}{2} \s W_{\a\b\g}~.
\eea
\end{subequations}

Let $\bar \F$ be a covariantly antichiral scalar, $\cD_\a \bar \F=0$, 
invariant under the super-Weyl transformations, $\d_\s \bar \F =0$. 
It is an instructive exercise to check, using the algebra of covariant derivatives given in the Appendix, 
the following super-Weyl  transformation law:
\bea
\d_\s \Big\{ \cD^2 \bar \cD^2 \bar \F + 8 \cD^\a (G_{\a\ad}\bar \cD^\ad \bar \F)\Big\}
&=& -(\s +\bar \s)  \Big\{ \cD^2 \bar \cD^2 \bar \F 
+ 8 \cD^\a (G_{\a\ad} \bar \cD^\ad\bar \F)\Big\}
\non \\
&& -2 \bar \cD_\ad \Big\{ (\bar \cD^\ad \bar \F) \cD^2 \s 
+ 4\ri (\cD^{\a\ad} \bar \F) \cD_\a \s \Big\} ~.
\label{second-order-operator}
\eea
To simplify the calculation, it is advantageous to make use of the identity
\bea
\d_\s \cD^2 = (\s-2\bar \s) \cD^2 + 2(\cD^\a \s) \cD_\a + \dots~,
\eea
where the ellipsis stands for all terms which involve the Lorentz generators 
and annihilate  scalars. 
It follows from \eqref{second-order-operator} that the operator $\D$ defined by
\bea
\D\bar \F := -\frac{1}{64} (\bar \cD^2 -4R ) \Big\{ \cD^2 \bar \cD^2 \bar \F 
+ 8 \cD^\a (G_{\a\ad}\bar \cD^\ad \bar \F)\Big\}~, \qquad 
\bar \cD_\ad \D\bar \F =0
\label{Delta}
\eea
has the super-Weyl transformation law
\bea
\d_\s \D\bar \F = -3\s \D\bar \F ~.
\label{sW-Delta}
\eea
We also introduce the conjugate operator 
\bea
\bar\D \F := -\frac{1}{64} ( \cD^2 -4\bar R ) \Big\{ \bar \cD^2  \cD^2  \F 
- 8 \bar \cD^\ad (G_{\a\ad} \cD^\a  \F)\Big\}~, \qquad 
 \cD_\a \bar \D  \F =0~.
\eea
For any covariantly chiral scalars $\F$ and $\J$, it holds that 
\bea
\int \rd^4 x \,\rd^2 \q \,\cE \, \J \D \bar \F &=& 
\int \rd^4 x \,\rd^2 \bar \q\, \bar \cE \, \bar \F \bar \D \J \non \\
&=& \frac{1}{16}\int \rd^4 x \,\rd^2 \q \,\rd^2 \bar \q \,E \,
 \Big\{ (\cD^2 \J) \bar \cD^2 \bar \F 
-8 (\cD^\a \J) G_{\a\ad}\bar \cD^\ad \bar \F\Big\} ~.~~~~
\label{symmetric}
\eea
In accordance with \eqref{sW-Delta},
this functional is super-Weyl invariant provided the chiral scalars $\F$ and $\J$
are inert under the super-Weyl transformations. 
In the case that $\F = \J$, the functional \eqref{symmetric} is real. 
Its component expression was given in \cite{FT} up to terms involving the gravitino.

The functional \eqref{symmetric} bears some resemblance to
\eqref{1.11} since both possess the same flat space limit, but the operators
underlying these two constructions are quite different. The chiral
d'Alembertian, which can be written
$\stackrel{\frown}{\Box}_+ = \frac{1}{16} (\bar \cD^2 - 4 R) \cD^2 + \cP$,
is a dimension-2 operator mapping a chiral multiplet to another chiral multiplet;
however, it cannot be made super-Weyl covariant for any choice of the chiral
function $\cP$, which was the point made in \cite{BK88}.
In contrast, the operator $\Delta$ underlying \eqref{symmetric}
is a dimension-3 super-Weyl covariant operator; it acts on
a weight-zero antichiral to yield a weight-three chiral superfield.
The overall chiral projector in \eqref{Delta} is critical for achieving
this \emph{manifest} super-Weyl covariance, because the quantity in
braces in \eqref{Delta} is neither super-Weyl covariant nor of definite
chirality by itself.
The resulting super-Weyl covariance of the functional \eqref{symmetric} is
absolutely critical for lifting the Fradkin-Tseytlin operator to superspace
and for enabling the construction which follows.

\section{Nonlocal effective action}

We now turn to deriving a nonlocal action that generates the super-Weyl anomaly \eqref{anomaly}.
We begin by introducing several important building blocks.

First, we require the super-Weyl transformations of some composite
expressions involving the torsion superfields. Using eq. \eqref{superweyl-torsion},
one can show
\bea
\d_\s (G^aG_a + 2R \bar R) &=& -(\s+\bar \s)  (G^aG_a + 2R \bar R) 
-\hf \bar \cD_\ad (\bar R \bar \cD^\ad \bar \s)  - \hf \cD^\a (R\cD_\a \s) \non \\ 
&&-\hf \bar \cD^\ad (G_{\a\ad}   \cD^\a \s)+\hf  \cD^\a (G_{\a\ad}  \bar  \cD^\ad \bar \s)~,
\eea
which guarantees that the functional
\bea
\int \rd^4 x \,\rd^2 \q \,\rd^2 \bar \q\,  E \,(G^aG_a + 2R \bar R) 
\eea
is super-Weyl invariant \cite{HT}.
Making use of the super-Weyl variation
\bea
\d_\s (\cD^2 R) = -(\s+\bar \s) \cD^2 R - 2\cD^\a(R\cD_\a \s) -\frac{1}{4} \cD^2\bar \cD^2 \bar \s~,
\eea
we further observe that 
\bea
\d_\s \Big\{ G^aG_a + 2R \bar R  &-& \frac{1}{4} \cD^2 R    \Big\}  = 
-(\s+\bar \s)  \Big\{ G^aG_a + 2R \bar R  - \frac{1}{4} \cD^2 R  \Big\}  \non \\
&+&\frac{1}{16} \cD^2\bar \cD^2 \bar \s
+\hf  \cD^\a (G_{\a\ad}  \bar  \cD^\ad \bar \s)
+\hf \bar \cD^\ad \Big( \bar R \bar \cD_\ad \bar \s
- G_{\a\ad}   \cD^\a \s\Big)~.
\eea
Now we introduce the composite chiral scalar
\bea 
\X:= -\frac{1}{4} (\bar \cD^2 -4R ) \Big\{ G^aG_a + 2R \bar R  
- \frac{1}{4} \cD^2 R \Big\}
\eea
and verify that
\bea
&& \d_\s \X = -3 \s \X + \D \bar \s~.
\eea

As a next step, we introduce two scalar Green's functions $G_{+-} (z, z') $ and $G_{-+} (z, z') $ 
that are related to each other by the rule 
\bea
G_{+-} (z, z') = G_{-+} (z', z)
\eea
and obey the following conditions: \\
(i)   the two-point function $G_{-+} (z, z') $ is covariantly antichiral
in $z$ and chiral in $z'$, 
\bea
\cD_\a  G_{-+} (z, z') =0~, \qquad \bar \cD_\ad' G_{-+} (z, z') =0~;
\eea
(ii) the two-point function $G_{-+} (z, z') $ satisfies the differential equation
\bea
\D G_{-+} (z, z') = \d_+ (z,z')~ .
\label{Green}
\eea
Here we have used the  chiral delta-function 
\bea
 \d_+ (z,z'):= -\frac{1}{4} (\bar \cD^2 -4R ) E^{-1} \d^4(x-x') 
 \d^2 (\q-\q') \d^2 (\bar \q -\bar \q') ~,
 \eea
which is covariantly chiral with respect to each of its arguments,
\bea
\bar \cD_\ad  \d_+ (z,z')=0~, \qquad \bar \cD'_\ad  \d_+ (z,z')=0~.
\eea
Under a finite super-Weyl transformation,  the delta-function 
$ \d_+ (z,z')$ can be seen to change as 
\bea
 \d_+ (z,z') \to \re^{-3\s}  \d_+ (z,z')~.
\label{sW-delta-function}
 \eea
It follows from the relations \eqref{sW-Delta}, \eqref{Green}
 and \eqref{sW-delta-function} that the Green's function $G_{-+} (z, z') $
is super-Weyl invariant, as is $G_{+-} (z, z') $.

Finally, it is useful to introduce a condensed notation
for the chiral and antichiral integration measures: 
\bea
\int \rd \m_+ := \int \rd^4 x \,\rd^2 \q \,\cE ~, \qquad
\int \rd \m_- := \int \rd^4 x \,\rd^2 \bar \q\, \bar \cE ~.
\eea

With the building blocks given above, let us consider two nonlocal real functionals 
\bea
\cF_1 &=&  \int \rd \m_+\int \rd \m'_-  \,\X(z) G_{+-} (z,z') \bar \X (z') ~,\\
\cF_2 &=&  \int \rd \m_+\int \rd \m'_- \, W^2(z) G_{+-} (z,z') \bar \X  (z')
~+ ~{\rm c.c.} ~,
\eea
with $W^2 = W^{\a\b\g}W_{\a\b\g} $.
The super-Weyl variations of these functionals are
\bea
\d_\s \cF_1 &=&  \int \rd \m_+ \, \s \X ~+ ~{\rm c.c.} ~, \\
\d_\s \cF_2 &=&  \int \rd \m_+ \, \s W^2 ~+ ~{\rm c.c.} 
\eea
We also note that the super-Weyl variation of the local functional \eqref{RbarR} is 
\bea
\d_\s \int \rd^4 x \,\rd^2 \q \,\rd^2 \bar \q \,E \, R \bar R = 
\frac{1}{16} \int \rd \m_+ \, \s 
(\bar \cD^2 -4R ) \cD^2 R ~+ ~{\rm c.c.} 
\label{3.18}
\eea
It follows from these relations that the  effective action can be chosen in the form:
\bea
8\p^2 \G &=&  (c-a)  \int \rd \m_+\int \rd \m'_-  \, W^2(z) 
G_{+-} (z,z') \bar \X  (z')   ~+ ~{\rm c.c.}\non \\
&-& a \int \rd \m_+\int \rd \m'_-  \, \X(z) 
 G_{+-} (z,z') \bar \X  (z')
+ (h+a) \int \rd^4 x \,\rd^2 \q \,\rd^2 \bar \q \,E \, R \bar R ~.~~~
\label{3.19}
\eea

\section{Concluding comments}

The nonlocal action \eqref{3.19} is one of the main results of our paper.
It provides the generalization of Riegert's construction\footnote{Several papers
\cite{EO,Deser96,Deser2000} pointed out certain pathological properties 
of the nonlocal action derived in \cite{Riegert}. More recently, this issue has been reconsidered in Ref.
\cite{Deser-rev} which ``rehabilitated'' \cite{Riegert}. Here we are only interested in the formal functional structure of the effective action.} 
 \cite{Riegert} to $\cN=1$ supergravity. 
The crucial property of $\G$ is that the corresponding supertrace $\langle T \rangle$, 
eq. \eqref{1.4}, coincides with the super-Weyl anomaly \eqref{anomaly}. 
The same anomaly is generated by the effective action derived twenty five 
years ago in \cite{BK88} (see eq. (14) in \cite{BK88}). 
The effective action derived in \cite{BK88}
differs from \eqref{3.19} by a super-Weyl invariant contribution 
that depends only on the gravitational superfield and therefore vanishes in any conformally flat superspace. 

In general, the super-Weyl anomaly may include not only the purely supergravity sector
\eqref{anomaly}, but also an additional contribution coming from a background vector supermultiplet, which is proportional to $\tr (W^\a W_\a)$, with $W_\a$ the covariantly chiral 
field strength. Such a contribution is taken into account by an additional correction 
to the effective action \eqref{3.19} obtained by the replacement 
\bea
W^{\a\b\g} W_{\a\b\g}~ \to ~\tr (W^\a W_\a)~.
\eea

The functional \eqref{symmetric} is super-Weyl invariant for arbitrary 
super-Weyl inert chiral scalars $\F$ and $\J$.  In a locally supersymmetric theory, 
these chiral scalars may be chosen to be some of the dynamical variables or composite objects. 
Thus \eqref{symmetric} generates a nontrivial higher derivative term in $\cN=1$ supergravity.  

Because the functional \eqref{symmetric} is super-Weyl invariant, it is independent of the chiral compensator $\vf$ and its conjugate.  Therefore, this functional can naturally be
described within any off-shell version of $\cN=1$ supergravity. 
As is known, any off-shell formulation for $\cN=1$ supergravity
can be realized by coupling a certain compensator to  
$\rm U(1)$ superspace \cite{Howe} (see \cite{GGRS}
for a review), which gauges not only the Lorentz
group but the $\rm U(1)$ $R$-symmetry group as well.\footnote{In its turn, U(1) superspace \cite{Howe} 
is a gauged-fixed version of  conformal superspace \cite{ButterN=1}.}
Using the same conventions
as in \cite{BK:Dual}, one finds that the operator $\Delta$ in $\rm U(1)$ superspace
is given by
\bea
\D\bar \F := -\frac{1}{64} (\bar \cD^2 -4R ) \Big\{ \cD^2 \bar \cD^2 \bar \F 
+ 8 \cD^\a (G_{\a\ad}\bar \cD^\ad \bar \F)
+ \frac{16}{3} \bar X_\ad \bar\cD^\ad \bar \F\Big\}~,
\quad \bar \cD_\ad \D\bar \F =0~,~~~~~
\label{DeltaU(1)}
\eea
and obeys the super-Weyl transformation 
\bea
\d_\L \D\bar \F = -3\L \D\bar \F ~,
\label{sW-DeltaU(1)}
\eea
where $\L$ is a real unconstrained super-Weyl transformation parameter.\footnote{We have flipped the
sign of $\L$ relative to \cite{BK:Dual}.} 
In U(1) superspace the functional 
\eqref{symmetric} turns into
\begin{align}
&\int \rd^4 x \,\rd^2 \q \,\cE \, \J \D \bar \F =
\int \rd^4 x \,\rd^2 \bar \q\, \bar \cE \, \bar \F \bar \D \J \non \\
& \quad =
\frac{1}{16}\int \rd^4 x \,\rd^2 \q \,\rd^2 \bar \q \,E \,
 \Big\{ (\cD^2 \J) \bar \cD^2 \bar \F 
-8 (\cD^\a \J) G_{\a\ad}\bar \cD^\ad \bar \F
-\frac{16}{3} (\bar\cD_\ad \bar X^\ad) \J \bar\F \Big\} ~.~~~~
\label{symmetricU(1)}
\end{align}
This functional
is real when $\J = \F$, keeping in mind that
$\bar\cD_\ad \bar X^\ad = \cD^\a X_\a$. We note that the superspace
formulation of new minimal supergravity
can be derived from $\rm U(1)$ superspace by taking $R=0$,
so the expressions \eqref{DeltaU(1)} and \eqref{symmetricU(1)} hold
equally there.

As we already emphasized, the functional \eqref{symmetric} is super-Weyl invariant
under the condition that $\F$ and $\J$ are super-Weyl inert chiral scalars. 
Choosing different types of superfields makes it possible to construct alternative 
superconformal invariants. For instance, if $U$ and $V$ are super-Weyl inert 
real scalars, then the functional 
\bea
 \frac{1}{16}\int \rd^4 x \,\rd^2 \q \,\rd^2 \bar \q \,E \, U \cD^\a(\bar \cD^2 -4R)\cD_\a V
\eea
is super-Weyl invariant. In the case that $U=V$ is a dynamical superfield,  
this functional coincides with the action of an Abelian vector multiplet \cite{WZ}. 
On the other hand,  $U$ and $V$ may be composite, for instance 
$U=V= K(\f^I, \bar \f^{\bar J}) $, where $K$ is the K\"ahler potential of a K\"ahler manifold, 
and $\f^I$ are super-Weyl inert chiral scalars. Then, the above functional generates 
a higher-derivative coupling which is invariant under arbitrary K\"ahler transformations 
$K(\f, \bar \f) \to K(\f , \bar \f)  +\L (\f) +\bar \L(\bar \f)$, as well as under arbitrary 
holomorphic isometries of the target K\"ahler  space.

The attempt to construct an $\cN=1$ supersymmetric extension 
of the Fradkin-Tseytlin 
operator was made a few years ago by Grosse \cite{Grosse}. 
Unfortunately, the attempt was unsuccessful\footnote{One of us (SMK)
had incorrectly argued to Johannes Grosse in 2006 and Yu Nakayama in 2012
that a superfield version of $\Delta_0$ could not exist, based
on the non-existence of the super-Weyl covariant
d'Alembertian $\stackrel{\frown}{\Box}_+$. While working on \cite{BdeWKL},
it became apparent that the higher dimension operator $\Delta$ could
play the required role in deriving  a supersymmetric extension of  Riegert's construction.}
and this author concluded
that ``there is no superfield version of the Riegert operator for chiral fields of 
Weyl weight 0.''
Although he did consider the integrand in the second line of
\eqref{symmetric}, he was not able to prove its super-Weyl invariance.
Most likely, this is due to its rather complicated transformation law.
We only discovered its invariance after first considering the associated
higher-order operator $\D$ defined by \eqref{Delta}, which proves to
possess the remarkably simple transformation law \eqref{sW-Delta}.
As demonstrated in \cite{BdeWKL}, the most natural origin of $\D$ is
within conformal superspace \cite{ButterN=1}.

In this paper we only considered the Wess-Zumino formulation for
$\cN=1$ supergravity \cite{WZ} which is characterized by  the old minimal set of auxiliary fields \cite{old}. 
As is well known, there exist two other off-shell formulations for $\cN=1$ supergravity,
the non-minimal \cite{non-min,SG} and the new minimal \cite{new}.
In the case of non-minimal  supergravity, the super-Weyl transformations were 
described by Siegel \cite{Siegel:SC}. For $n=-1$ non-minimal supergravity, 
a superfield Fradkin-Tseytlin operator was constructed in \cite{Manvelyan}. 
Super-Weyl transformations and trace anomalies in various off-shell versions
for $\cN=1$ supergravity have recently been discussed in \cite{Bonora}.
\\

\noindent
{\bf Acknowledgements:}\\
We thank Arkady Tseytlin for bringing the important references 
\cite{FT1982} to our attention. 
SMK is grateful to Stefan Theisen for a useful discussion and for pointing Ref. \cite{Deser-rev} out. 
The work of DB was supported by
ERC Advanced Grant No. 246974, ``{\it Supersymmetry: a window to
non-perturbative physics}.''
The work of SMK   was supported in part by the Australian Research Council,
project No. DP1096372.

\appendix

\section{The Wess-Zumino superspace geometry}

In describing the Wess-Zumino superspace geometry (see \cite{WB} for a review), 
we mostly follow the notation and conventions of \cite{BK}.\footnote{These
conventions are nearly identical to those
of Wess and Bagger \cite{WB}. To convert the notation of \cite{BK} to that of \cite{WB}, one
replaces $R \rightarrow 2 R$, $G_{\alpha \ad} \rightarrow 2 G_{\alpha \ad}$, and
$W_{\alpha \beta \gamma} \rightarrow 2 W_{\alpha \beta \gamma}$.
In addition, the vector derivative has to be changed by the rule 
$\cD_a \to \cD_a +\frac{1}{4}\ve_{abcd}G^b M^{cd}$, where $G_a$ corresponds to \cite{BK}. 
Finally, the spinor Lorentz generators $(\s_{ab})_\a{}^\b$ and 
$({\tilde \s}_{ab})^\ad{}_\bd$  used in \cite{BK} have an extra minus sign as compared with \cite{WB}, 
specifically $\s_{ab} = -\frac{1}{4} (\s_a \tilde{\s}_b - \s_b \tilde{\s}_a)$ and 
 $\tilde{\s}_{ab} = -\frac{1}{4} (\tilde{\s}_a {\s}_b - \tilde{\s}_b {\s}_a)$.  
 Unlike \cite{BK}, in this paper the full superspace measure is denoted 
$E= {\rm Ber} \,(E_M{}^A)$. } 
In particular,  the coordinates of $\cN=1$ curved superspace $\cM$ are denoted 
$z^M = (x^m, \q^\m , {\bar \q}_{\dot \m})$.
The superspace geometry is described 
by covariant derivatives of the form
\bea
\cD_A &=& (\cD_a , \cD_\a ,\cDB^\ad ) = E_A + \O_A~.
\eea
Here $E_A$ denotes the inverse vielbein, 
$E_A = E_A{}^M  \pa_M $,
and $\O_A$  the Lorentz connection, 
\bea
\O_A = \hf\,\O_A{}^{bc}  M_{bc}
= \O_A{}^{\b \g} M_{\b \g}
+\O_A{}^{\bd \gd} {\bar M}_{\bd \gd} ~,
\eea
with $M_{bc} \Leftrightarrow ( M_{\b\g}, {\bar M}_{\bd \gd})$
the Lorentz generators.
The covariant derivatives obey the following anti-commutation relations:
\begin{subequations}\label{algebra}
\bea
\{ \cD_\a , {\bar \cD}_\ad \} &=& -2{\rm i} \cD_{\a \ad} ~, \qquad
\{\cD_\a, \cD_\b \} = -4{\bar R} M_{\a \b}~, \qquad
\{ {\bar \cD}_\ad, {\bar \cD}_\bd \} =  4R {\bar M}_{\ad \bd}~,  \\
\left[ { \bar \cD}_{\ad} , \cD_{ \b \bd } \right]
     & = & -{\rm i}{\ve}_{\ad \bd}
\Big(R\,\cD_\b + G_\b{}^{\dot{\g}}  \cDB_{\dot{\g}}
-(\cDB^\gd G_\b{}^{\dot{\d}})
{\bar M}_{\gd \dot{\d}}
+2W_\b{}^{\g \d}
M_{\g \d} \Big)
- {\rm i} (\cD_\b R)  {\bar M}_{\ad \bd}~,~~~~~~~  \\
\left[ \cD_{\a} , \cD_{ \b \bd } \right]
     & = &
     {\rm i}
     {\ve}_{\a \b}
\Big({\bar R}\,\cDB_\bd + G^\g{}_\bd \cD_\g
- (\cD^\g G^\d{}_\bd)  M_{\g \d}
+2{\bar W}_\bd{}^{\gd \dot{\d}}
{\bar M}_{\gd \dot{\d} }  \Big)
+ {\rm i} (\cDB_\bd {\bar R})  M_{\a \b}~.
\label{algebra_d}
\eea
\end{subequations}
The torsion tensors $R$, $G_a = {\bar G}_a$ and
$W_{\a \b \g} = W_{(\a \b\g)}$ satisfy the Bianchi identities
\begin{subequations}
\bea
\cDB_\ad R&=& 0~, \qquad \cDB_\ad W_{\a \b \g} = 0~, \label{2.4a} \\
\cDB^\gd G_{\a \gd} &=& \cD_\a R~, \label{2.4b} \\
\cD^\g W_{\a \b \g} &=& {\rm i} \,\cD_{(\a }{}^\gd G_{\b) \gd}~.
\label{2.4c} 
\eea
\end{subequations}

\begin{footnotesize}

\end{footnotesize}

\end{document}